# Theoretical study of the magnetization dynamics of non-dilute ferrofluids

D.V. Berkov[1], L.Yu. Iskakova[2], A.Yu. Zubarev[2]

[1]*Innovent Technology Development, Pruessingstr. 27B. D-07745, Jena, Germany*

[2]*Ural State University, Lenina Ave 51, 620083, Ekaterinburg, Russia*

## ABSTRACT

The paper is devoted to the theoretical investigation of the magnetodipolar interparticle interaction effect on remagnetization dynamics in moderately concentrated ferrofluids. We consider a homogenous (without particle aggregates) ferrofluid consisting of identical spherical particles and employ a rigid dipole model, where magnetic moment of a particle is fixed with respect to the particle itself. In particular, for the magnetization relaxation after the external field is instantly switched off, we show that the magnetodipolar interaction leads to the increase of the initial magnetization relaxation time. For the complex *ac*-susceptibility $\chi(\omega) = \chi'(\omega) + i\chi''(\omega)$ we find that the this interaction leads to an overall increase of $\chi''(\omega)$ and shifts the $\chi''(\omega)$ - peak towards lower frequencies. Comparing results obtained with our analytical approach (second order virial expansion) to numerical simulation data (Langevin dynamics method), we demonstrate that the employed virial expansion approximation gives a good qualitative description of the ferrofluid magnetization dynamics and provides a satisfactory quantitative agreement with numerical simulations for the *dc* magnetization relaxation - up to the particle volume fraction $\phi \sim 10\%$ and for the ac-susceptibility - up to $\phi \approx 5$ %.

## I. INTRODUCTION

In this paper we study the remagnetization dynamics in ferrofluids - colloidally stable suspensions of magnetic single-domain particles in a carrier liquid (in order to prevent aggregation of particles due to the magnetodipolar attraction, ferrofluid particles are covered with the special surfactant layers). Due to the possibility to change physical parameters and control the behavior of a ferrofluid by an externally applied field, such systems are of a large interest both for fundamental and applied physics. Ferrofluids are used in many existing technologies and are supposed to be highly promising for a variety of potential technical and medical applications [1].

Experiments demonstrate that magnetodipolar interparticle interaction changes significantly both the equilibrium [2] and dynamical [3] properties of ferrofluids. Theoretical models of dynamical properties of dilute ferrofluids with vanishing interparticle interactions have been proposed in [4, 5, 6, 7]. These models lead to very accurate results for very dilute ferrofluids but can not explain properties and behavior of ferrofluids where the interparticle interaction is significant.

Depending on the energy of this magnetodipolar interaction, it can lead either to an appearance of homogeneous short- and long-range interparticle correlations, or to a formation of chain-like, drop-like and other heterogeneous internal structures [8]. At present there is no general theory allowing to predict internal structure in non-dilute ferrofluids at given experimental conditions. Therefore it is reasonable to consider effects of different internal structures on the dynamical phenomena in ferrofluids separately. Such idealized models can provide better insights into the influence of various structures and factors on the macroscopical properties of ferrofluids.

Combination of corresponding idealized models can serve as a basis for constructing theories of real magnetic fluids with typical long-range interparticle correlations, where also various heterogeneous particle aggregates are present.

However, before considering dynamical properties of a ferrofluid with various particle aggregates, the behavior of a homogeneous system should be properly understood. For this reason, in this paper we present a model of the remagnetization dynamics of a homogeneous ferrofluid consisting of identical particles. It is assumed that magnetic moment of each particle has a constant magnitude and is "frozen" into the particle body. We realize, that this model is obviously the oversimplification of real ferrofluids. The problem is not only a more or less broad distribution of geometric and magnetic particle parameters of real ferrofluids (this can be easily included into the approaches used by us) and particle aggregates often present in real systems (such aggregates obviously require a special treatment). A very important physical aspect also is, that the intrinsic magnetic anisotropy of an individual ferrofluid particle is finite (and usually not even large compared to the thermal energy and magnetodipolar interaction field), so that the magnetic moment can rotate with respect to the particle itself. Still, the analysis of a simple model studied here is a mandatory first step for understanding dynamical properties of real ferrofluids.

Remagnetization dynamics of the model outlined above is studied both analytically and using numerical simulations. To obtain analytical results with maximal mathematical accuracy, we take into account the magnetodipolar interparticle interaction using the regular method of virial expansion over the particle concentration. We assume that magnetodipolar interaction energy is less or of the same order of magnitude as the thermal energy $kT$. Otherwise particle aggregates must appear in a ferrofluid, which treatment is out of the framework of this paper. Next, in order to focus on the effects of magnetodipolar interaction, we neglect here effects of the hydrodynamical interaction between particles. Effects of this interaction will be considered in a separate publication.

The paper is organized in the following way. In the next Section we explain in detail our analytical approach, deriving the governing equation for the macroscopical magnetization dynamics. In Sec. III we present the numerical simulation methodology and justify our the choice of the short-range repulsive potential. In Sec. IV we study the effect of the magnetodipolar interparticle interaction first, on the magnetization relaxation after a step-wise (instant) change of the external field and second, on the *ac*-susceptibility of a ferrofluid. Here we calculate corresponding dynamical system behavior and compare results of the analytical approach to numerical simulation studies, establishing the concentration region where the analytical theory provides a quantitatively accurate description of the magnetization dynamics in ferrofluids.

## II. ANALYTICAL APPROACH AND BASIC EQUATION FOR THE MAGNETIZATION DYNAMICS

We consider a ferrofluid with volume $V$ containing $N$ identical spherical ferromagnetic particles with the diameter $d$. The absolute magnitude $p_{\text{mag}}$ of the particle magnetic moment $\mathbf{p}_{i,\text{mag}}$ is constant, the moment is "frozen" into the particle body. We introduce the unit vector $\mathbf{m}_i = \mathbf{p}_{i,\text{mag}} / p_{\text{mag}}$ of the magnetic moment of the $i$-th particle and denote the particle radius-vector by $\mathbf{r}_i$.

To calculate the macroscopic characteristics of this system, we must determine the $N$-particle distribution function $P_N(\mathbf{m}_1,\ldots\mathbf{m}_N,\mathbf{r}_1,\ldots\mathbf{r}_N)$. It can be found by solving the appropriate Fokker-Planck equation, where we have to take into account magnetodipolar interactions between all particles. This corresponding equation is:

$$\frac{\partial P_N}{\partial t} = \sum_i \mathbf{I}_i \left( \frac{D_\text{r}}{kT} \mathbf{I}_i \left( U P_N \right) \right) + \sum_i \nabla_i \left( \left( \frac{D_\text{t}}{kT} \nabla_i U \right) P_N \right) + D_\text{r} \sum_i \mathbf{I}_i^2 P_N + D_\text{t} \sum_i \nabla_i^2 P_N \qquad (1)$$

where the summation in (1) is performed over particles and we have used the standard notation $\mathbf{I}_i = \mathbf{m}_i \times \frac{\partial}{\partial \mathbf{m}_i}, \quad \nabla_i = \frac{\partial}{\partial \mathbf{r}_i}$.

The potential energy of the system

$$U = -kT\sum_i (\boldsymbol{\kappa}\cdot\mathbf{m}_i) + \frac{1}{2}\sum_{i\neq j} w_{ij}$$

contains the energy due to the external field (first term), where the reduced field $\boldsymbol{\kappa} = \mu_0 \frac{p_{\text{mag}}\mathbf{H}}{kT}$ is defined via the vacuum permeability $\mu_0$, and the local magnetic field $\mathbf{H}$.

Magnetodipolar interaction energy (2[nd] term in the expression for $U$) contains pair interaction terms $w_{ij}$ (interaction energy of particles $i$ and $j$)

$$w_{ij} = \frac{\mu_0}{4\pi} p_{\text{mag}}^2 \frac{(\mathbf{m}_i\cdot\mathbf{m}_j) r_{ij}^2 - 3(\mathbf{m}_i\mathbf{r}_{ij})(\mathbf{m}_j\mathbf{r}_{ij})}{r_{ij}^5}$$

where $\mathbf{r}_{ij}$ is the radius-vector between the centers of these particles.

Rotational and translational particle diffusion coefficients

$$D_\text{r} = \frac{kT}{6V_\text{p}\eta}, \quad D_\text{t} = \frac{kT}{3\pi\eta d},$$

are determined by the hydrodynamical (including the non-magnetic shell) particle diameter $d$, the carrier fluid viscosity $\eta$ and the particle volume $V_\text{p} = \pi d^3/6$. In solving Eq. (1) we must take into account the condition that the particles can not overlap: $r_{ij} \geq d$.

The Fokker-Planck equation (1) can not be solved exactly for two reasons. The first one is the well-known problem of statistical physics – interparticle interaction in a many-particle system does not allow (nearly always) to solve the governing equation for a many-particle distribution function or to calculate the Gibbs statistical integral. The second reason is the purely mathematical difficulty arising by the solution of the Fokker-Planck equation even for the single particle.

In order to overcome the second problem, we use the effective-field approach, suggested in [4], which is a version of the trial function method. According to this approach, we write the function $P_N$ in the form of an equilibrium Gibbs function in some effective magnetic field $\mathbf{H}_\text{e}$, which must be determined, instead of the real field $\mathbf{H}$. With other words, we postulate the validity of the following equation:

$$\sum_i \mathbf{I}_i\left(\frac{D_\text{r}}{kT}\mathbf{I}_i(U_\text{e}P_N)\right) + \sum_i \nabla_i\left(\left(\frac{D_\text{t}}{kT}\nabla_i U_\text{e}\right)P_N\right) + D_\text{r}\sum_i \mathbf{I}_i^2 P_N + D_\text{t}\sum_i \nabla_i^2 P_N = 0 \tag{2}$$

where

$$U_\text{e} = -kT\sum_i(\boldsymbol{\kappa}_\text{e}\cdot\mathbf{m}_i) + \frac{1}{2}\sum_{i\neq j} w_{ij}, \qquad \boldsymbol{\kappa}_\text{e} = \mu_0\frac{p_{\text{mag}}\mathbf{H}_\text{e}}{kT}$$

Combining (1) and (2), we obtain:

$$\frac{\partial P_N}{\partial t} = \sum_i \mathbf{I}_i\left(\frac{D_\text{r}}{kT}\mathbf{I}_i(\delta\boldsymbol{\kappa}P_N)\right), \quad \delta\boldsymbol{\kappa} = \boldsymbol{\kappa}_\text{e} - \boldsymbol{\kappa} \tag{3}$$

Up to this point all transformations were exact: instead of the unknown function $P_N$ we have introduced the unknown reduced field $\boldsymbol{\kappa}_\text{e}$ linked to $P_N$ by the Gibbs formula

$$P_N = Z^{-1}\exp\left(-\frac{U_\mathrm{e}}{kT}\right), \quad Z = \int \exp\left(-\frac{U_\mathrm{e}}{kT}\right) d\mathbf{m}_1 ... d\mathbf{m}_N d\mathbf{r}_1 ... d\mathbf{r}_N \tag{4}$$

The crucial assumption of this method is that the effective field does not depend on the vectors $\mathbf{m}_i$ and $\mathbf{r}_i$ and that the components of this field can be found from the equation for the first statistical moment of the function $P_N$ . Note that for dilute ferrofluids this method leads to a very good agreement with experiments and results of computer simulations (see, e.g., [9]). Similar ideas have been successfully used by analyzing rheological properties of ferrofluids with chain-like aggregates [10].

As usual in statistical physics, in a general case interparticle interactions do not allow to calculate exactly the average values of physical quantities using Eq. (4). From now on we suppose that particle concentration in our system is not high and use the virial expansion method. In Sec. V we show that this method represents a reasonable approximation when describing dynamical properties of low and moderately concentrated ferrofluids.

It is convenient, first, to average the distribution function $P_N$ over coordinates $\mathbf{r}_i$ of all particles. Introducing the Mayer function $f_{ij}$ = exp($-w_{ij}/kT$)-1 and averaging (4) over all $\mathbf{r}_i$, we obtain the averaged $N$-particle distribution function $p_N$ in the form

$$p_N = \int P_N \prod_i d\mathbf{r}_i = Z^{-1}\exp\left(\boldsymbol{\kappa}_\mathrm{e}\cdot\sum_l \mathbf{m}_l\right)\int\prod_{i>j}\left(1+f_{ij}\right)\prod_k d\mathbf{r}_k \tag{5}$$

Expanding (5) in a power series in $f_{ij}$, keeping only the first two terms and performing standard transformations, we obtain

$$p_N = \left(\prod_k \psi_k\right)\left(1-\phi\left(N-1\right)G_\mathrm{e} + \frac{1}{V}\sum_{i>j} Q_{ij}\right) \tag{6}$$

Here $\phi = NV_\mathrm{p}/V$ is the hydrodynamical (including the non-magnetic particle shell) volume concentration of particles and

$$\psi_i = \psi(\mathbf{m}_i) = \frac{\exp(\boldsymbol{\kappa}_\mathrm{e}\cdot\mathbf{m}_i)}{z_1}, \quad z_1 = \int \exp(\boldsymbol{\kappa}_\mathrm{e}\cdot\mathbf{m})d\mathbf{m} = 4\pi\frac{\sinh\kappa_\mathrm{e}}{\kappa_\mathrm{e}},$$

$$Q_{ij} = \int_{r_{ij}} f_{ij} d\mathbf{r}_{ij}, \quad G_\mathrm{e} = \frac{1}{2V_\mathrm{p}}\langle\psi_1\psi_2 Q_{12}\rangle_{12}, \quad \langle ...\rangle_{i...k} = \int ... \prod_{l=i}^{k} d\mathbf{m}_l \qquad i,k = 1,...N$$

Calculating the integral in the expression for $Q_{ij}$, we must keep in mind that the result depends on the shape of the (infinite) integration volume [10]. The reason for this is the long-range character of the dipolar interaction. A physically correct way of integration must provide for a system in thermodynamical equilibrium the equality of the magnetic field in the integration volume to the local physical field $\mathbf{H}$ in the sample region where interacting particles are situated. For this reason we must use the integration cave as an infinitely long cylinder, directed along the field $\mathbf{H}$, with the trial (first) particle on the axis of this cylinder, integrating over all positions of the 2$^{\mathrm{nd}}$ particle. Technically this means that in the integral for $Q_{ij}$ (see above) we should use a cylindrical coordinate system ($\rho$, $\varphi$, $z$) with the $z$-axis along $\mathbf{H}$. First, we must integrate over the $z$-coordinate of the 2$^{\mathrm{nd}}$ particle (from $-\infty$ to $+\infty$, then over other coordinates. This integration order has been successfully used in [11] to calculate the ferrofluid equilibrium magnetization.

Unfortunately, the complicated form of the Mayer function makes the analytical calculation of $Q_{ij}$ impossible. Here we restrict ourselves to the situation when the dipolar interaction energy $w_{ij}$ between particles is $w_{ij} \sim kT$ or smaller. Obviously this assumption means that there are no heteroaggregates in the system.

Expanding the Mayer function in a power series in $w_{ij,}$, keeping only the linear terms and using the method from [11] to calculate of the integral for $Q_{ij}$, we obtain

$$Q_{ij} = 8\gamma\left(\mathbf{m}_i \cdot \mathbf{m}_j\right), \quad G(x) = 4\lambda L^2(x), \qquad L(x) = \coth x - \frac{1}{x}, \quad G_\mathrm{e} = G(\kappa_\mathrm{e}) \tag{7}$$

where the interaction parameter $\lambda = \dfrac{\mu_0}{4\pi}\dfrac{p_\mathrm{mag}^2}{d^3 kT}$ characterizes the ratio of the dipolar interaction energy of two closely placed particles to the thermal energy $kT$.

Averaging Eq. (3) over the particle positions, we come to an equation identical to (3) with $p_N$ instead of $P_N$. Multiplying the resulting equation by $\mathbf{m}_1$ and averaging over all $\mathbf{m}_i$, we obtain

$$\frac{\partial \boldsymbol{\mu}}{\partial t} = -D_\mathrm{r}\left\langle \mathbf{m}_1 \sum_i \mathbf{I}_i\left(\left[\delta\boldsymbol{\kappa}\times\mathbf{m}_i\right] p_N\right)\right\rangle, \text{ with } \langle\ldots\rangle \;= \langle\ldots\rangle_{1\ldots N} \tag{8}$$

Here $\boldsymbol{\mu} = \langle \mathbf{m}_1 p_N \rangle$ is the average of the orientation vector $\mathbf{m}_1$ of the trial particle. The ferrofluid magnetization is

$$\mathbf{M} = n p_\mathrm{mag}\boldsymbol{\mu}, \qquad n = \frac{N}{V} = \frac{\phi}{V_\mathrm{p}} \tag{9}$$

Using Eqs. (6) and (7) we find (introducing $L_\mathrm{e} = L(\kappa_\mathrm{e})$)

$$\boldsymbol{\mu}(\kappa) = \mu_\mathrm{e}\mathbf{e}_h, \qquad \mu_\mathrm{e} = L_\mathrm{e} + \frac{N-1}{V}V_\mathrm{p}\frac{\mathrm{d}G_\mathrm{e}}{\mathrm{d}\kappa_\mathrm{e}}, \qquad \mathbf{e}_h = \frac{\mathbf{H}_\mathrm{e}}{H_\mathrm{e}}, \tag{10}$$

In the thermodynamical limit (prime means a derivative with respect to $\kappa_\mathrm{e}$)

$$\mu_\mathrm{e} = L_\mathrm{e} + \phi\frac{dG_\mathrm{e}}{d\kappa_\mathrm{e}} = L_\mathrm{e} + 8\phi\lambda L_\mathrm{e} L_\mathrm{e}' \tag{11}$$

Taking into account that the angular momentum operator $\mathbf{I}$ is antihermitian and using the approximation (6) for $p_N$, we obtain:

$$\left\langle \mathbf{m}_1 \sum_i \mathbf{I}_i\left(\left[\delta\boldsymbol{\kappa}\times\mathbf{m}_i\right] p_N\right)\right\rangle = \langle \boldsymbol{\xi} p_N\rangle = \langle \boldsymbol{\xi}\psi_1\rangle + \frac{N-1}{V}\mathrm{v}\left[\langle \boldsymbol{\xi} b\psi_1\rangle_1 - 2G_\mathrm{e}\langle \boldsymbol{\xi}\psi_1\rangle_1\right] \tag{12}$$

where $\boldsymbol{\xi} = \left[\mathbf{m}_1 \times \left[\delta\boldsymbol{\kappa}\times\mathbf{m}_1\right]\right]$ and $b = \langle \psi_2 Q_{12}\rangle_2$, and in the thermodynamical limit

$$\left\langle \mathbf{m}_1 \sum_i \mathbf{I}_i\left(\left[\delta\boldsymbol{\kappa}\times\mathbf{m}_i\right] p_N\right)\right\rangle = \langle \boldsymbol{\xi} p_N\rangle = \langle \boldsymbol{\xi}\psi_1\rangle + \phi\left[\langle \boldsymbol{\xi} b\psi_1\rangle_1 - 2G_\mathrm{e}\langle \boldsymbol{\xi}\psi_1\rangle_1\right] \tag{13}$$

Combining Eqs.(6), (7) and (13), after simple transformations we have

$$\left\langle \mathbf{m}_1 \sum_i \mathbf{I}_i\left(\left[\delta\boldsymbol{\kappa}\times\mathbf{m}_i\right] p_N\right)\right\rangle = A_\mathrm{e}\delta\boldsymbol{\kappa} - B_\mathrm{e}(\mathbf{e}_h\cdot\delta\boldsymbol{\kappa})\mathbf{e}_h \tag{14}$$

where the functions

$$A_\mathrm{e}=A(\kappa_\mathrm{e}), \quad B_\mathrm{e}=B(\kappa_\mathrm{e}) \tag{15}$$

are defined via

$$A(x) = 1 - L(x)/x + 8\phi\lambda\cdot(L^2(x) - C(x))\cdot L(x)/x$$

$$B(x) = C(x) + 8\varphi\lambda \cdot (L^2(x) - C(x)) \cdot L(x)/x, \qquad C(x) = 1 - 3L(x)/x$$

Substituting Eq. (14) into (8) and (10), we obtain

$$\frac{\partial \boldsymbol{\mu}}{\partial t} = -D_{\mathrm{r}} \left[ A_{\mathrm{e}} \delta \boldsymbol{\kappa} - B_{\mathrm{e}} \left( \mathbf{e}_h \cdot \delta \boldsymbol{\kappa} \right) \mathbf{e}_h \right] \tag{16}$$

We arrived at a system of equations (10),(11) and (16) for the vectors $\boldsymbol{\mu}$ and $\boldsymbol{\kappa}_{\mathrm{e}}$. For subsequent calculations it is convenient to write this system in the form of a single equation with respect to $\boldsymbol{\kappa}_{\mathrm{e}}$. To this end we employ the fact that $\boldsymbol{\mu} = \mu_{\mathrm{e}} \mathbf{e}_h$ and write

$$\frac{\partial \boldsymbol{\mu}}{\partial t} = J_{\mathrm{e}} \mathbf{e}_h \frac{\partial \kappa_{\mathrm{e}}}{\partial t} + \mu_{\mathrm{e}} \frac{\partial \mathbf{e}_h}{\partial t}, \tag{17}$$

$$J_{\mathrm{e}} = \frac{d\mu_{\mathrm{e}}}{d\kappa_{\mathrm{e}}} = L'_{\mathrm{e}} + 8\varphi\lambda \left( (L')^2 + L_{\mathrm{e}} L''_{\mathrm{e}} \right)$$

Substituting (17) into the first equation of (16) and writing the scalar product of the result and vector $\mathbf{e}_h$, we obtain

$$\frac{d\kappa_{\mathrm{e}}}{dt} = \frac{D_{\mathrm{r}}}{J_{\mathrm{e}}} \left( B_{\mathrm{e}} - A_{\mathrm{e}} \right) \left( \mathbf{e}_h \cdot \delta \boldsymbol{\kappa} \right) \tag{18}$$

Finally, inserting (19) into (17) and the result into (16), we arrive at the equation

$$\frac{d\mathbf{e}_h}{dt} = -D \frac{A_{\mathrm{e}}}{\mu_{\mathrm{e}}} \left( \delta \boldsymbol{\kappa} - \mathbf{e}_h \left( \mathbf{e}_h \cdot \delta \boldsymbol{\kappa} \right) \right) \tag{19}$$

Equations (18) and (19) form a system of equations for $\kappa_{\mathrm{e}}$ and $\mathbf{e}_h$, which can be easily reduced to a single equation

$$\frac{d\boldsymbol{\kappa}_{\mathrm{e}}}{dt} = -D_{\mathrm{r}} \left[ \frac{A_{\mathrm{e}} - B_{\mathrm{e}}}{\kappa_{\mathrm{e}}^2 J_{\mathrm{e}}} \left( \boldsymbol{\kappa}_{\mathrm{e}} \delta \boldsymbol{\kappa} \right) \boldsymbol{\kappa}_{\mathrm{e}} + \frac{A_{\mathrm{e}}}{\mu_{\mathrm{e}}} \kappa_{\mathrm{e}} \left( \delta \boldsymbol{\kappa} - \frac{\left( \boldsymbol{\kappa}_{\mathrm{e}} \cdot \delta \boldsymbol{\kappa} \right)}{\kappa_{\mathrm{e}}} \boldsymbol{\kappa}_{\mathrm{e}} \right) \right] \tag{20}$$

To find the macroscopical magnetization $\mathbf{M} = np_{\mathrm{mag}}\boldsymbol{\mu}$, we have to solve Eq. (20) and substitute the result into (10).

## III. NUMERICAL SIMULATIONS METHODOLOGY

Our numerical simulations are based on the Langevin dynamics formalism, where the equations of motion for the relevant degrees of freedom characterizing our system are solved taking into account thermal fluctuations.

For the ferrofluid model considered in this paper (particles with ‘fixed’ magnetic moments, which are not allowed to move with respect to the particles itself) the system of equations for the description of ferrofluid dynamics includes two equations - for the translational and rotational particle motions. For the time scale of interest (~ $10^{-6}$ sec) inertial terms can be neglected due to small particle sizes (~ 10 nm) and a substantial carrier fluid viscosity (~ 0.1 Ps) typical for ‘standard’ ferrofluids.

In this approximation the equation for the translational particle motion in a ferrofluid can be simply deduced from the balance between the viscous force $-b \cdot d\mathbf{r}/dt$ and all other forces acting on the $i$-th ferrofluid particle:

$$b_i \frac{d\mathbf{r}_i}{dt} = \nabla(\mathbf{p}_{i,\mathrm{mag}} \mathbf{H}_i^{\mathrm{dip}}) - \nabla U_i^{\mathrm{rep}} + \mathbf{F}_i^{\mathrm{fl}} \tag{21}$$

Here $b$ denotes the viscous friction coefficient, which for a spherical particle with the hydrodynamical radius $R_{\mathrm{hyd}}$ in a fluid with the viscosity $\eta$ is $b = 6\pi\eta R_{\mathrm{hyd}}$. The first term on the right-hand side represents the magnetodipolar interaction force $\mathbf{F}^{\text{dip}} = -\nabla U^{\text{dip}} = \nabla(\mathbf{p}_{\text{mag}}\mathbf{H}^{\text{dip}})$ and the second one - the steric repulsion force $\mathbf{F}^{\text{rep}} = -\nabla U^{\text{rep}}$. This latter force is due to the non-magnetic shell surrounding the magnetic particle kernel. The choice of the repulsive potential $U^{\text{rep}}$ will be discussed in detail below. The third term is a stochastic thermal force $\mathbf{F}^{\text{fl}}$ responsible for a translational Brownian motion. This force has $\delta$-functional correlation properties

$$\langle F^{\text{fl}}_{i,\xi}(0) \cdot F^{\text{fl}}_{j,\psi}(t)\rangle = 2kTb_i \cdot \delta_{ij}\delta_{\xi\psi}\delta(t)$$

in our model, where the hydrodynamic interaction between particles is neglected.

Employing the same approximations, we can write the equation for particle rotational motion as the balance between the viscous torque and all the other torques:

$$\zeta_i \frac{d\mathbf{p}_{i,\text{mag}}}{dt} = -\mathbf{p}_{i,\text{mag}} \times \left[\mathbf{p}_{i,\text{mag}} \times \mathbf{H}^{\text{dip}}_i\right] - \left[\mathbf{p}_{i,\text{mag}} \times \mathbf{T}^{\text{fl}}_i\right] \tag{22}$$

Here $\zeta_i = 8\pi\eta R_{\text{hyd}}{}^3$ is the rotational viscous friction coefficient. The first term on the r.h.s. is the torque exerted on the magnetic moment by the magnetodipolar interaction field $\mathbf{H}^{\text{dip}}$. This torque is directly 'transferred' on the particle itself due to the 'fixed moment' approximation of our model. The random torque $\mathbf{T}^{\text{fl}}$ due to the thermal bath fluctuations leads in the Langevin dynamics formalism to the rotational Brownian motion of the particle. If the hydrodynamic interaction is be neglected, the components of $\mathbf{T}^{\text{fl}}$ have the same simple correlation properties as for the random force $\mathbf{F}^{\text{fl}}$: $\langle T^{\text{fl}}_{i,\xi}(t) \cdot T^{\text{fl}}_{j,\psi}(t')\rangle = 2kT\zeta_i \cdot \delta_{ij}\delta_{\xi\psi}\delta(t-t')$.

The system of stochastic differential equations (SDE) (21)-(22) is solved by the optimized Bulirsch-Stoer method (see [12] for the description of the basic idea of this algorithm), which converges to the Stratonovich solution of these SDEs. Methods for the numerical evaluation of the long-range magnetodipolar field $\mathbf{H}^{\text{dip}}$ present in both equations (21) and (22) are discussed in detail in our review [13]. For this study, where a formation of particle aggregates was not expected (the absence of such aggregates was confirmed by simulations), we have chosen the modified Lorentz cavity method. We have used the cut-off radius of the Lorentz sphere $R_{\text{L}} = 2\langle\Delta r\rangle$, where $\langle\Delta r\rangle$ is the mean interparticle distance. It was checked that further increase of $R_{\text{L}}$ does not affect simulation results within statistical errors.

All results presented below have been obtained for a system of particles with the magnetic core radius $R_{\text{mag}}$ = 6 nm, non-magnetic shell thickness $h$ = 2 nm and the particle material magnetization M = 400 G, for which the interaction parameter $\lambda$ (defined after Eq. (7)) is $\lambda = 0.8$.

The next important methodical question is the choice of the short-range repulsive potential $U^{\text{rep}}$ present in (21). The corresponding issue was discussed in [13] from the 'physical' point of view, i.e., considering the plausibility of the choice for $U^{\text{rep}}$ as a 'representative' for a steric repulsion force acting between surfactant-coated magnetic particles in real ferrofluid. In this particular research, however, we have an additional methodical problem: taking into account that one of the main goals of this study is the comparison between analytical theory and numerical simulations, we have to choose the repulsive potential in such a way that it does introduce an artificial bias into such a comparison.

The simplest choice which would enable a most straightforward comparison between analytical theory and numerical simulations would be the hard-core potential ($U^{\text{rep}} = 0$ for $\Delta r > 2R_{\text{hyd}}$ and $U^{\text{rep}} = \infty$ for $\Delta r < 2R_{\text{hyd}}$). This choice would exactly correspond to the condition that particles are not allowed to overlap used by the analytical solution of the basic Eq. (1). Unfortunately, the hard core-potential is not differentiable, so that the dynamic equation containing it can not be solved in a standard way. Instead, the so called 'collision-based' algorithms (see [14] for the review of these methods) should be employed, where the evaluation of the next collision time is used to determine the maximal time step and the system behavior after the particle collision. Such algorithm performs quite well when the hard-core potential only exists in the system under study. However, in the presence of another potential (like the magnetodipolar interaction present in our case), the evaluation of the collision time becomes a delicate matter and the 'collision-based' algorithms are known to work very slow.

For this reason we have chosen several kinds of analytical short-range potentials and tested whether and how the simulation results depend on the kind of $U^{\text{rep}}$. Our first choice was the purely exponential (Yukawa-type) potential $U_1^{\text{rep}}(r) = B\exp(-(r-2R_{\text{mag}})/h)$. The decay radius of this potential is equal to the non-magnetic shell thickness $h$ and the amplitude $B$ is chosen to be much larger than the maximal magnetodipolar interaction energy of particles with the interparticle distance equal to the magnetic core diameter: $E_{\text{max}}^{\text{dip}} = (\pi/3)M^2V_{\text{mag}}$ (here $M$ is the particle magnetization and $V_{\text{mag}}$ - the volume of the magnetic particle core).

The second potential tested by us was the potential of the screened-Coulomb type

$$U_2^{\text{rep}}(r) = A_q \frac{\exp(-s/q)}{s}, \qquad s = \frac{r-2R_{\text{mag}}}{h} \tag{23}$$

Here the constant $q$ controls the screening radius $r_{\text{scr}} = hq$, and the amplitude $A_q$ was chosen so that the repulsion force due to the potential (23) was equal to the maximal magnetodipolar attraction force acting between particles placed at the distance $\Delta r = 2R_{\text{hyd}}$. When the constant q decreases, the screening radius $r_{\text{scr}} \to 0$, and the amplitude $A_q \to \infty$ preserving the property that the repulsion force $F_{\text{rep}}(\Delta r = 2R_{\text{hyd}})$ is equal to the maximal magnetodipolar attraction force. In this sense we can say that this repulsive potential converges to the hard-core potential with $R_{\text{core}} = 2R_{\text{hyd}}$ when $q \to 0$.

Test simulation results for the exponential potential $U_1^{\text{rep}}(r) = B\exp(-(r-2R_{\text{mag}})/h)$ with $B = 10\cdot E_{\text{max}}^{\text{dip}}$ and screened-Coulomb potentials $U_2^{\text{rep}}(r)$ with two very different values of the constant $q$ ($q$ = 0.5 and $q$ = 4.0) are shown in Fig. 1. Potential dependencies on the interparticle distance for all three potentials are shown in Fig. 1(a). The magnetization time-dependencies $m(t)$, computed for these three types of $U^{\text{rep}}$ after the initially applied magnetic field $H$ = 200 Oe is instantly switched off, are displayed in the part (b). One can see that within the statistical simulation errors all the time-dependencies for all three potentials fully coincide, thus ensuring the independence of the simulation results on the choice of the short-range potential for our system. This proves that the differences between the analytical theory and numerical simulations observed and discussed below are not due to an improper choice of the short-range repulsive potential in our simulations.

Concluding this discussion, we remind that the question concerning the dependence of the *equilibrium* ferrofluid behavior on the exact form of short-range repulsion potential was studied analytically in [15] (see also references therein). The main result of this study was that for dilute and moderately concentrated ferrofluid where the three-particles correlations are not very important, the equilibrium magnetization of a homogeneous (without particle aggregates) ferrofluid *does not depend* on the form of this short-range potential. Our numerical simulations show that this conclusion remains true also for the *dynamical* properties of a ferrofluid.

## IV. RESULTS AND DISCUSSION

### IV. 1. Remagnetization dynamics after an instantaneous change of an applied field

Let us assume that at the time $t$ = 0 the magnitude of an applied field changes instantaneously from the initial value $H_1$ to the final one $H_2$, whereby direction of the field remains the same.

The *analytical approach* outlined above (Sec. II) leads in this case to the following version of the Eq. (20):

$$\frac{d\kappa_e}{dt} = -D_r \left[ \frac{A_e - B_e}{J_e} \left( \kappa_e - \kappa_2 \right) \right] \tag{24}$$

where the initial condition is

$$\kappa_e = \kappa_1 \quad \text{at} \quad t = 0 \tag{25}$$

with $\kappa_{1,2} = \mu_0 p_{mag} H_{1,2} / kT$ .

The Cauchy problem (24,25) can be easily solved with any commercially available software package capable to handle ordinary differential equations.

*Numerical simulations* of the remagnetization dynamics are performed in the following way. We start with the system of particles which magnetic moments are aligned in the direction of the external field $\mathbf{H}_1$. The system is equilibrated in this field until the magnetization does not changed anymore (in frames of statistical errors); the 'annealing' time interval $\Delta t_{ann} = 5t_{Br}$ ($t_{Br}$ is the Brownian relaxation time) is usually long enough to achieve this equilibrium. Afterwards, the external field is instantly changed to $\mathbf{H}_1$ and the magnetization relaxation is recorded. To achieve a high accuracy required, in particular, to determine the relaxation time, we have performed the averaging over $N_{att} = 32$ independent runs for a system of $N_p = 1000$ particles.

Corresponding analytical and numerical simulation results are compared for the step-wise decrease and increase of the applied field in Fig. 2 and 4.

For the magnetization decay after the external field is switched off (Fig. 2), one can see that the analytical model agrees with the simulation results fairly well for ferrofluids with concentration of the magnetic phase up to $\phi \approx 6$ %, what represents - from the "applied" point of view - a moderately concentrated ferrofluid. It is interesting to note that the substantial contribution to the disagreement between analytical theory and numerical simulations results from the corresponding disagreement between the initial (equilibrium) magnetization values. The latter is due to the overestimation of the equilibrium ferrofluid magnetization by the second order virial expansion approach.

This important issue is illustrated in Fig. 3, which shows the concentration dependence of the relaxation time defined as

$$t_{rel} = \left| \frac{\langle m_z \rangle}{d\langle m_z \rangle / dt} \right|_{\kappa_e = \kappa_1} \tag{26}$$

after the instant field decrease from $H_1 = 200$ Oe to $H_2 = 0$, which corresponds to the initial stage of the relaxation process shown in Fig. 2.

This plot demonstrates, on the one hand, that the magnetic interaction between particles increases the magnetization relaxation time (decreases its relaxation rate) at the initial relaxation stage. In the studied concentration range the increase of $t_{rel}$ is nearly linear with concentration. This increase is caused by the formation of short-range correlation between particle moments; the corresponding correlation degree increases with the particle concentration due to the magneto-dipolar interparticle interaction.

On the other hand, Fig. 3 shows that a good agreement between the analytical model and numerical simulations concerning the *initial* relaxation time persists up to the highest particle volume concentration studied here ($\phi = 14$ %, which from the experimental point of view means a highly concentrated ferrofluid), so that the analytical theory predicts this dynamical system feature much better than its equilibrium magnetization value.

The same line of arguments allows to explain why the agreement between theory and simulations is much better (persists up to higher concentrations) when the external field is initially absent ($H(t = 0) = 0$) and then is instantly switched on (see Fig. 4). In this case, first of all, the initial (for $t = 0$) equilibrium magnetization is, of course, absent ($\mathbf{M}(t = 0) = 0$) both in analytical theory and simulations. Moreover, one can show analytically that in the second order virial expansion the initial slope of the magnetization curve $dm_z(t)/dt$ does not depend on the particle concentration. Using numerical simulations we have verified, that this analytical result is valid up to the highest studied concentration $\phi = 14\%$. So for the magnetization *increase* after the external field is switched *on*, the discrepancy between analytical theory and simulation results arises due to the different rate of the magnetization change when the system becomes magnetized up to some extent, as it can be seen from Fig. 4.

Concluding this subsection, we would like to consider the magnetization relaxation after an instant change of the external field when the effective field $\kappa_e$ is *nearly equal* to the final field $\kappa_2$ ($|\kappa_e - \kappa_2| / \kappa_2 \ll 1$). In this case the relaxation time $\tau_2$, which characterizes this 'linear' remagnetization dynamics, exhibits a non-trivial dependence on the final field value $\kappa_2$, as we show below.

In the linear approximation with respect to $\delta\kappa = \kappa_e - \kappa_2$ equation (24) can be written as

$$\frac{d\kappa_e}{dt} = -D_r \left[ \frac{A_2 - B_2}{J_2} (\kappa_e - \kappa_2) \right], \qquad A_2, B_2, J_2 = A, B, J(\kappa_2) \tag{27}$$

One can easily show that in the same approximation Eq. (27) leads to

$$\frac{d\mu}{dt} = -\frac{\mu - \mu_2}{\tau_2}, \text{ where } \tau_2 = \frac{1}{D_r} \left[ \frac{J_2}{A_2 - B_2} \right], \quad \mu_2 = \mu(\kappa_2) \tag{28}$$

which allows a straightforward calculation of the relaxation time $\tau_2$.

Corresponding results presenting the relaxation time $\tau_2$ as a function of the final field $\kappa_2$ are shown in Fig. 5. One can see that the interaction between particles increases $\tau_2$ (i.e. decreases the remagnetization rate) when the field $\kappa_2$ is relatively weak and decreases $\tau_2$ (accelerates the remagnetization) when $\kappa_2$ is high.

Such a non-trivial dependence of $\tau_2$ on the final field $\kappa_2$ is a result of the competition between two factors. The first one is the usual effect of the interparticle interaction, which decreases the relaxation rate analogous to the remagnetization dynamics after a large change of an external field occurs (see above). The second factor is the well known effect of the increase of a mean particle magnetic moment due to the interaction between particles [9, 11]. The last factor increases the remagnetization rate. When the final field is weak or moderate, the first factor dominates, when the field is strong – the second one.

### IV.2. Complex susceptibility

In this section we study the ferrofluid response to a linearly polarized oscillating field

$$H_z = H_0 \cdot \sin\omega t, \quad H_x = H_y = 0 \tag{29}$$

*Analytical approach*. Equation (20) now reads

$$\frac{d\kappa_e}{dt} = -D_r \left[ \frac{A_e - B_e}{J_e} (\kappa_e - \kappa_0 \sin\omega t) \right], \quad \kappa_0 = \mu_0 \frac{p_{\mathrm{mag}} H_0}{kT} \tag{30}$$

This equation can be also easily solved numerically. Substituting $\kappa_e(t)$ obtained from (30) into Eqs. (10) and (11), we find the mean $z$-projection of the moment unit vector $\mu_z(t) \equiv \langle m_z(t) \rangle$. The Fourier transforms

$$\mu'(\Omega) = \int_0^\infty \mu_z(t)\sin(\Omega t)dt \qquad\qquad \mu''(\Omega) = \int_0^\infty \mu_z(t)\cos(\Omega t)dt \tag{31}$$

provide the real $\mu'(\Omega)$ and imaginary $\mu''(\Omega)$ parts of $\mu_\omega$, related to the corresponding parts of the ferrofluid magnetization Fourier transform as $M_\omega = np_{mag}\mu_\omega$. The index $\omega$ means here that the applied field oscillates with the frequency $\omega$.

We define the reduced complex susceptibility

$$\chi_{red}(\omega) = \frac{\mu_\omega(\omega)}{h_0} , \tag{32}$$

where the reduced field is defined via the saturation magnetization of the particle material $M_S$ as $h_0 = H_0/M_S$. The reduced susceptibility (32) is proportional to the standard susceptibility $\chi = M/H_0$ which describes the reaction of the ferrofluid at the same frequency $\omega$ as the frequency of the applied field. However, the reduced quantity $\chi_{red}$ is more convenient to study the effects of the interparticle interaction, because the trivial proportionality of the standard susceptibility $\chi = M/H_0$ to the particle concentration is eliminated (we remind that $\mu_z(t)$ is the average $z$-projection of the magnetic moment unit vector).

*Numerical 'measurements'* of the complex susceptibility are straightforward and described in detail in our review [13]. In short, we start simulations from the state with chaotically oriented particle magnetic moments and 'anneal' the system during $\Delta t_{ann} = t_{Br}$ in the absence of an external field. A shorter annealing time - compared to the simulations of the magnetization relaxation described above - is possible, because the average magnetization does not change during the equilibration process, so that only short-range correlations between the particle moments have to be established. Afterwards, we switch on the oscillating field $\mathbf{H} = H_0\mathbf{e}_z\sin(\omega t)$ and compute the in-phase and out-of-phase responses of the $z$-component of magnetization ($L$ is the number of the time steps). Dividing the results by the field amplitude and by the saturation magnetization of the system (in order to eliminate the proportionality of $\chi = M/H$ to the particle concentration, as by the definition (32))

$$\begin{aligned} \mathrm{Re}(\chi_{red}) \equiv \chi'_{red} &= \frac{1}{h_0}\cdot\frac{1}{L}\sum_{l=1}^{L}\langle m_z(t_l)\rangle\cdot\sin(\omega t_l) \\ \mathrm{Im}(\chi_{red}) \equiv \chi''_{red} &= \frac{1}{h_0}\cdot\frac{1}{L}\sum_{l=1}^{L}\langle m_z(t_l)\rangle\cdot\cos(\omega t_l) \end{aligned} \tag{33}$$

we obtain the complex susceptibility *per particle* $\chi_{red}$. To obtain the frequency dependence of the *ac*-susceptibility at a given temperature $\chi_{red}(\omega)$, we perform the 'measurements' (33) at a set of frequencies sufficiently 'dense' to resolve all features of this dependence. To obtain the results with a sufficiently 'good' statistics, we have carried out the simulations during $N_{cyc} = 5$ field cycles at each frequency (so that simulations are especially time-consuming in the low-frequency region), and performed the averaging over $N_{att} = 8$ independent runs for a system with $N_p = 500$ particles each.

Fig. 6 demonstrates the comparison of analytical results obtained using (30-32) and numerical simulations for the real $\chi'_{red}$ and imaginary $\chi''_{red}$ susceptibility parts. From the qualitative point of view, both analytical approach and numerical simulations predict the shift of the peak on the imaginary susceptibility part ($\chi''_{red}(\omega)$-dependence) towards *lower* frequencies with increasing particle volume fraction $\phi$. This is in a qualitative agreement with the increase of the relaxation time $t_{rel}$ with the growing particle concentration discussed above. Quantitatively, we note that the disagreement between analytical theory and numerical simulations is more significant (for one and same particle concentration) than for the magnetization relaxation study performed in the previous subsection. The explanation of this phenomenon can be as follows. For all concentrations, the deviation between the analytical approach and simulation results for the imaginary part of the *ac*-susceptibility has *different* signs for low and high frequencies (see Fig. 6). Taking into account, that the magnetization relaxation after an instantaneous change of an external field contains contributions from *all* frequencies, the difference between the 'analytical' and 'numerical' susceptibilities may be partially 'averaged out' for the magnetization relaxation process.

## CONCLUSION

In this paper we have studied the influence of the magnetodipolar interparticle interaction on the remagnetization dynamics of a homogeneous ferrofluid using an analytical model and numerical simulations. The analytical model is based on the regular second order virial approximation and does not contain any adjustable parameters or heuristic constructions. It leads to a good quantitative agreement with computer simulation results (which can be considered as exact for our ferrofluid model) up to the volume concentration of magnetic phase $\phi \sim$ 5-10%, depending on the type of the remagnetization dynamics under study. We note, that these volume concentration can be considered as being relatively high from the point of view of modern ferrofluid applications.

Our results show, that the magnetodipolar interaction increases the characteristic time of the magnetization decay immediately after the applied field is switched off. For the magnetization relaxation for the case when the initial field is close to the final one, the relaxation time demonstrates a more complicated behavior, increasing with the particle concentration if the final field is weak and decreasing if this field is strong. The main effect of the magnetodipolar interaction on the frequency dependence of the ferrofluid *ac*-susceptibility is twofold: this interaction enhances its imaginary part, and shifts the peak on the $\chi''(\omega)$-dependence towards lower frequencies, in accordance with the increase of the system relaxation time mentioned above.

Our study of the ferrofluid dynamics has been performed for the 'fixed dipole' model, where the particle magnetic moment is fixed with respect to the particle itself. The understanding of this simple model is the necessary first step for the theoretical analysis of this complex system. However, we point out, that in order to properly understand the behavior of real ferrofluids, the inclusion of the hydrodynamical interparticle interaction and the extension of the model to allow for the internal magnetic degrees of freedom (rotation of the magnetic moment relative to the particle due to the finite value of the single-particle magnetic anisotropy) is necessary.

**Acknowledgements**

This work has been done under the financial support of RFFI, grants N 06-01-00125, 07-02-00079, 07-01-960769Ural, Fund CRDF, PG07-005-02.

**Figure captions**

Fig. 1. Magnetization relaxation curves after switching the applied field $H$ = 200 off (at $t/t_{visc}$ = 100) for a ferrofluid with the magnetic particle volume fraction $\phi$ = 10 % simulated with various short-range repulsion potentials $U(r)$ as shown in the legends. It can be clearly seen that results for various $U(r)$ coincide within statistical errors.

Fig. 2. Comparison of the analytical theory (open circles) and numerical simulation results (solid lines): magnetization relaxation $m_z(t)$ after switching off the external field $H$ = 200 Oe at $t$ = 0 for various volume fractions of magnetic particles $\phi$. Analytical results agree reasonably well with numerical simulation up to the concentration $\phi \approx 6\%$. Note that the disagreement between simulations and analytics is largely due to the difference between the initial (equilibrium) magnetization values $m_{eq}(H = 200)$. Particle parameters: magnetic core radius $R_p$ = 6 nm, shell thickness $h$ = 2 nm, magnetization of the core material $M$ = 400 G.

Fig. 3. Concentration dependence of the initial relaxation time calculated analytically using the definition (26) (solid line) and computed numerically from the simulated relaxation curves $m_z(t)$ as described in the paper text (open squares, dashed line is a guide for an eye). In contrast to relaxation curves, 'analytical' and 'numerical' initial relaxation times nearly agree (within statistical errors of numerical simulations) up to the highest studied concentration $\phi$ = 14 %.

Fig. 4. The same as in Fig. 2 for the dc-magnetization, when the field $H$ = 200 Oe is instantly switched on at $t$ = 0. Particle parameters are the same as in Fig. 2. Note, that the agreement between analytical results and numerical simulations is much better than for the $m_z(t)$-relaxation after switching the external field off (compare to the Fig. 2).

Fig. 5. Dependence of the magnetization relaxation time $t_{rel}$ after the instant change of the applied field $H_{init} \to H_{fin}$ on its final value $H_{fin}$ when the initial field $H_{init}$ is only slightly smaller than $H_{fin}$ for various particle concentrations as shown in the legend. Note that the relaxation time $t_{rel}$ increases with the particle concentration $\phi$ for small final fields $H_{fin}$ < 20, but decreases with $\phi$ for large fields $H_{fin}$ > 20.

Fig. 6. Real (top series) and imaginary (bottom series) parts of the complex susceptibility $\chi(\omega)$ of a ferrofluid with the same particle parameters as on previous figures for three particle concentrations as indicated in the legend. Note, that the agreement between numerical results (full squares) and analytical values (open circles) for the susceptibility is significantly worse then for the magnetization relaxation (compare curves on e.g., Fig. 2 and on this figure for one and the same particle concentration).

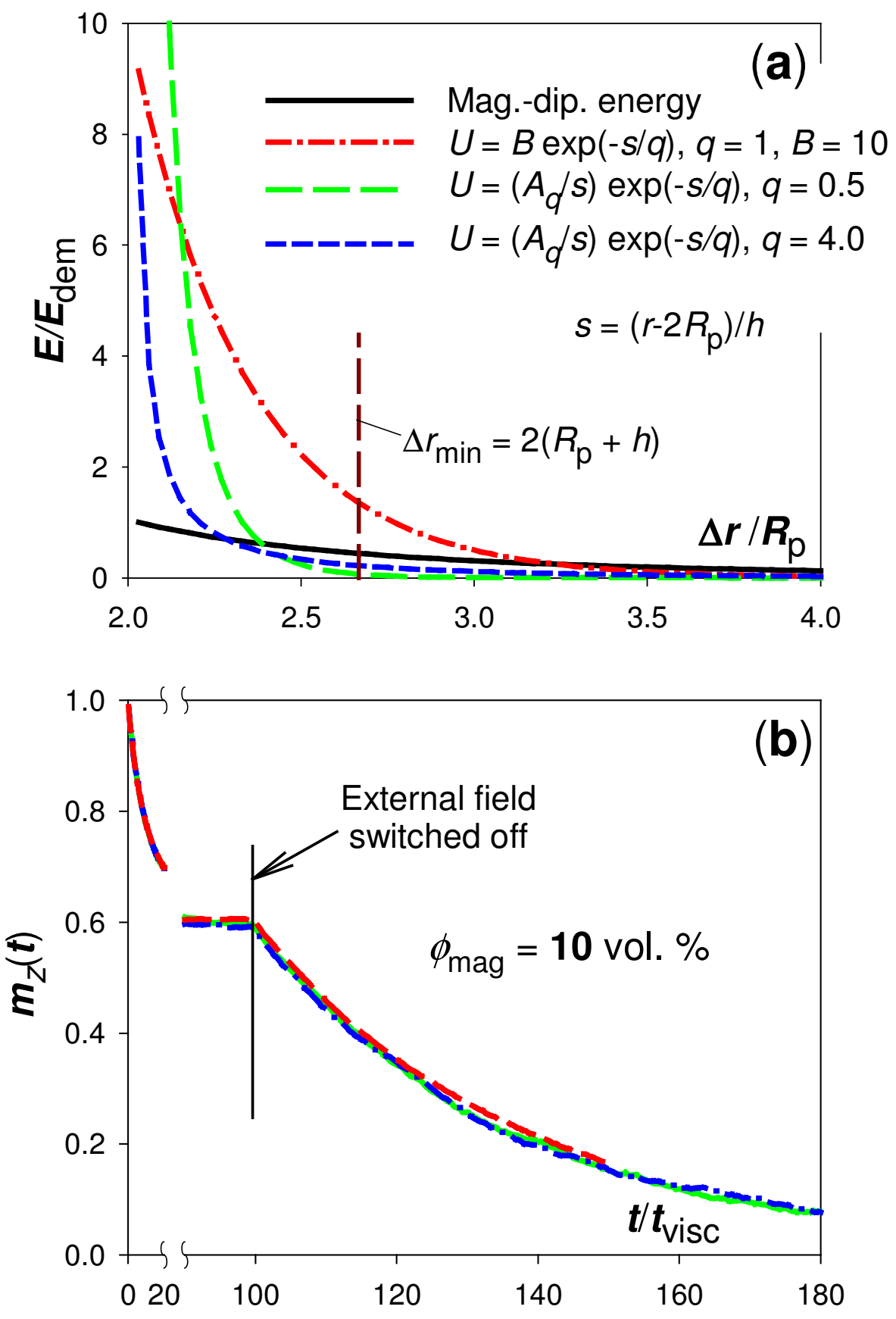
(a)
Mag.-dip. energy
U = B exp(-s/q), q = 1, B = 10
U = (A_q/s) exp(-s/q), q = 0.5
U = (A_q/s) exp(-s/q), q = 4.0
s = (r-2R_p)/h
Δr_min = 2(R_p + h)
E/E_dem
Δr / R_p
(b)
External field switched off
φ_mag = 10 vol. %
m_z(t)
t/t_visc

Fig. 1

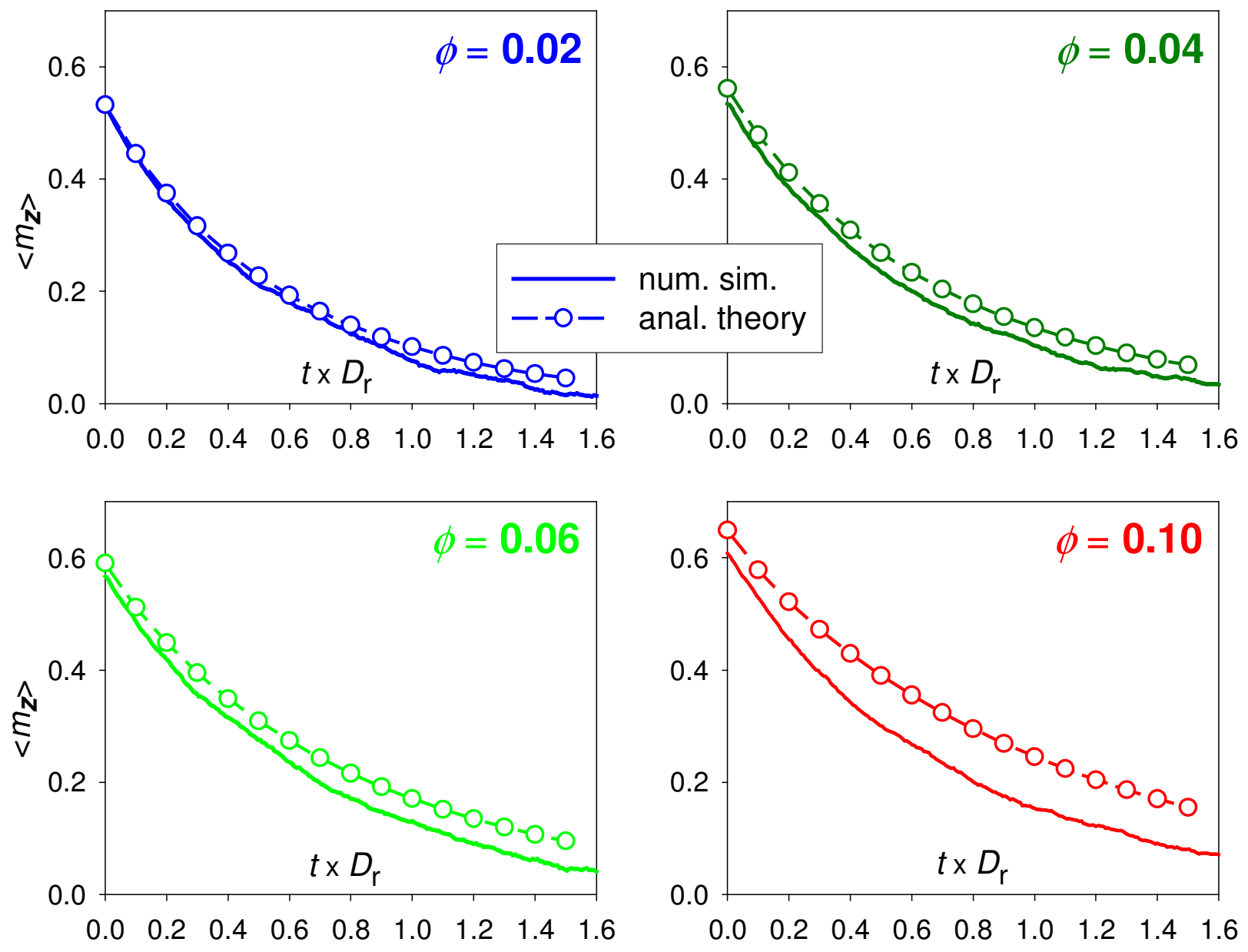

Fig. 2

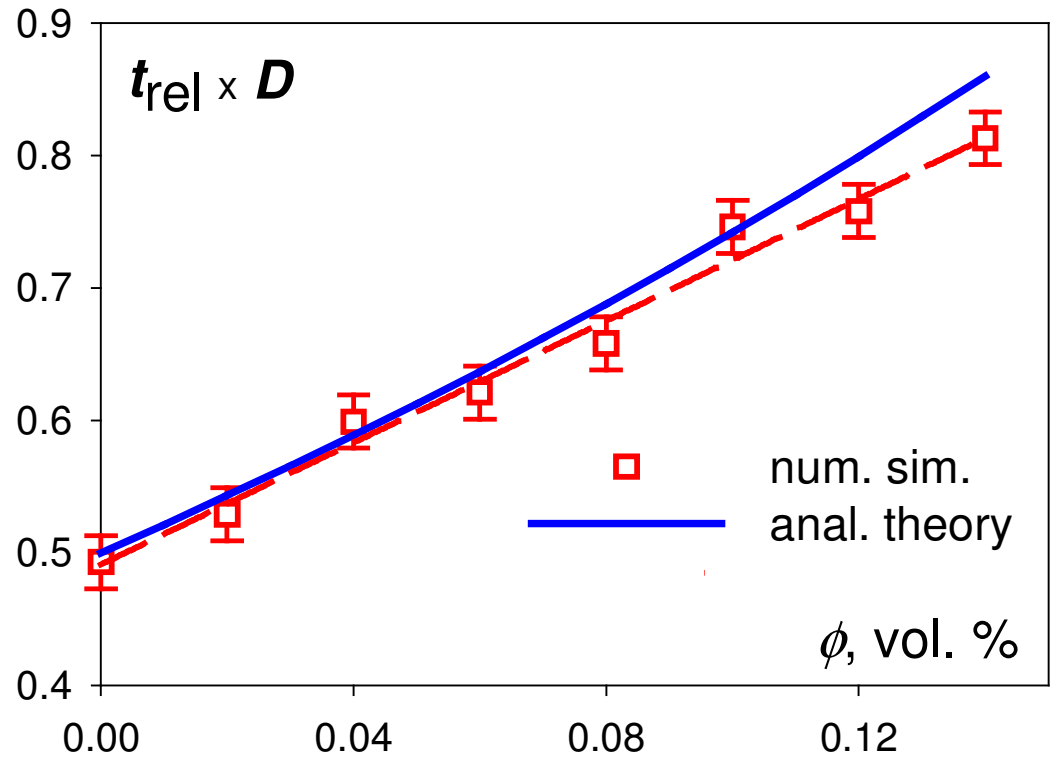

Fig. 3

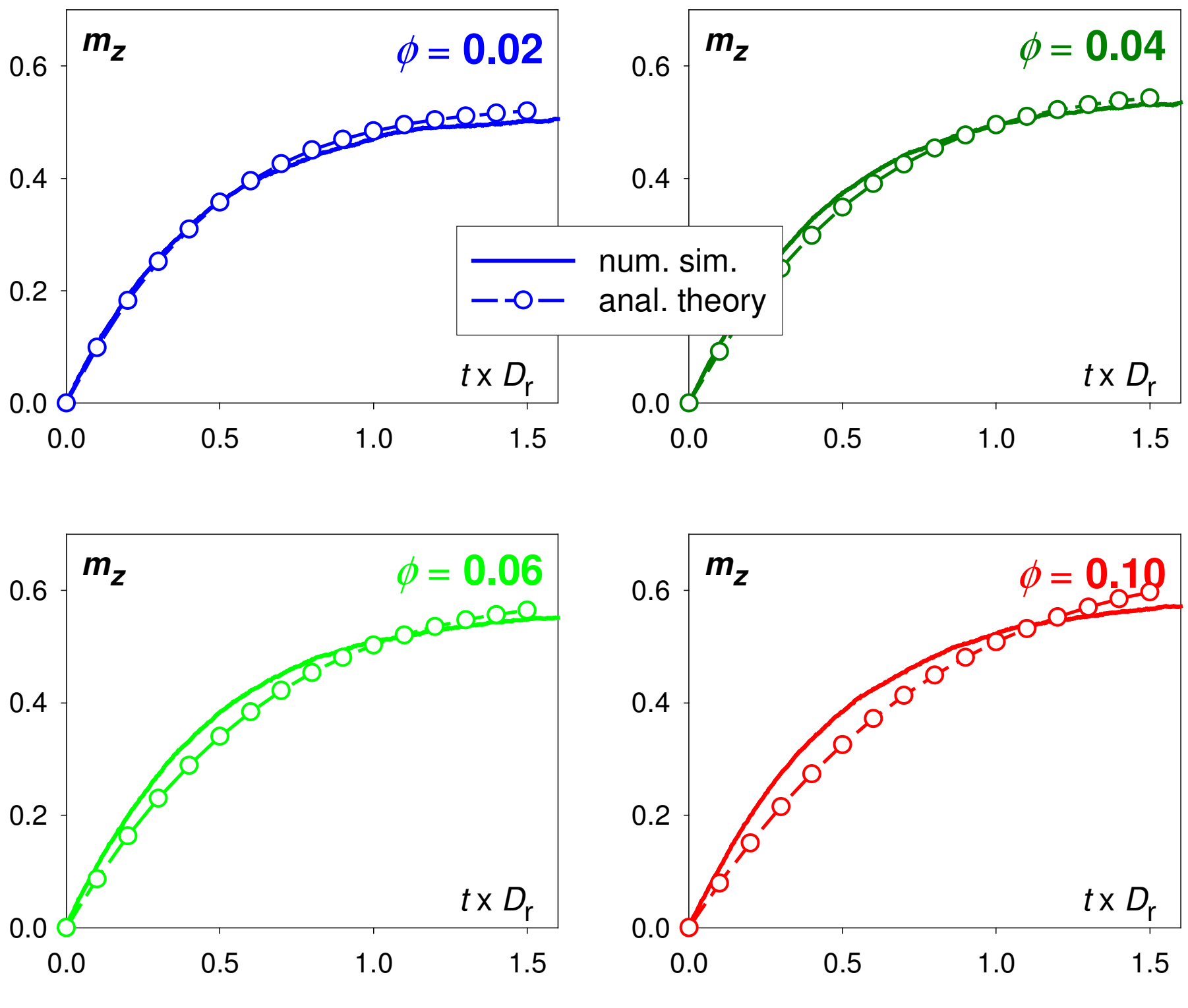

Fig. 4

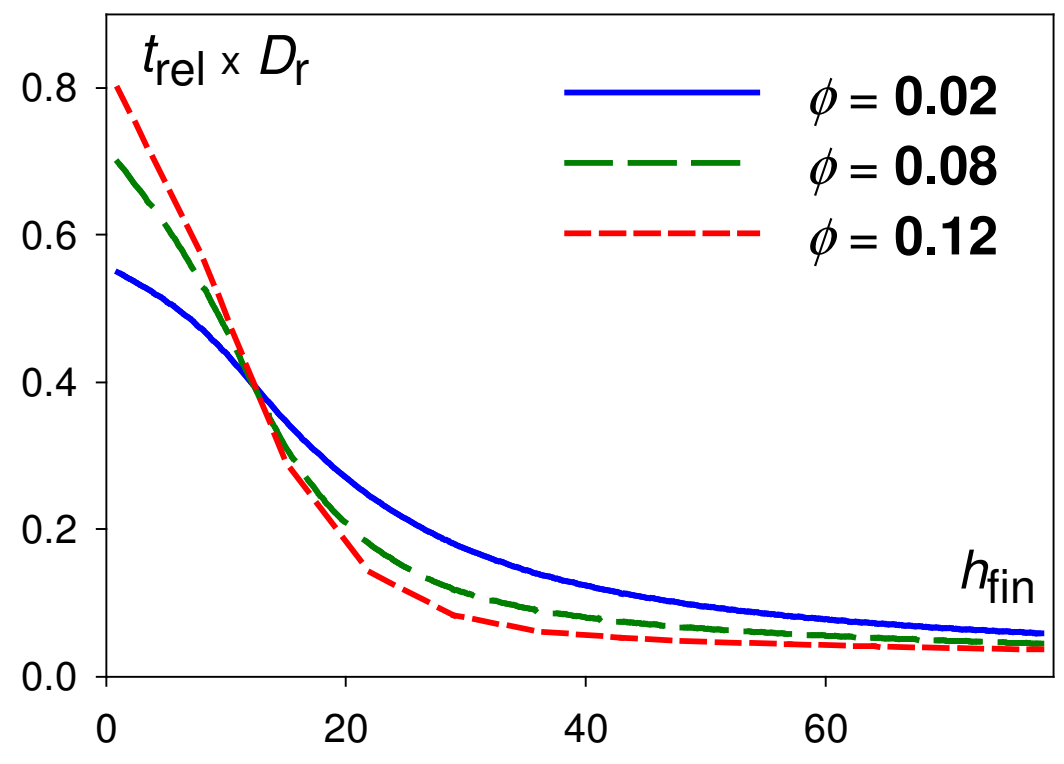

Fig. 5

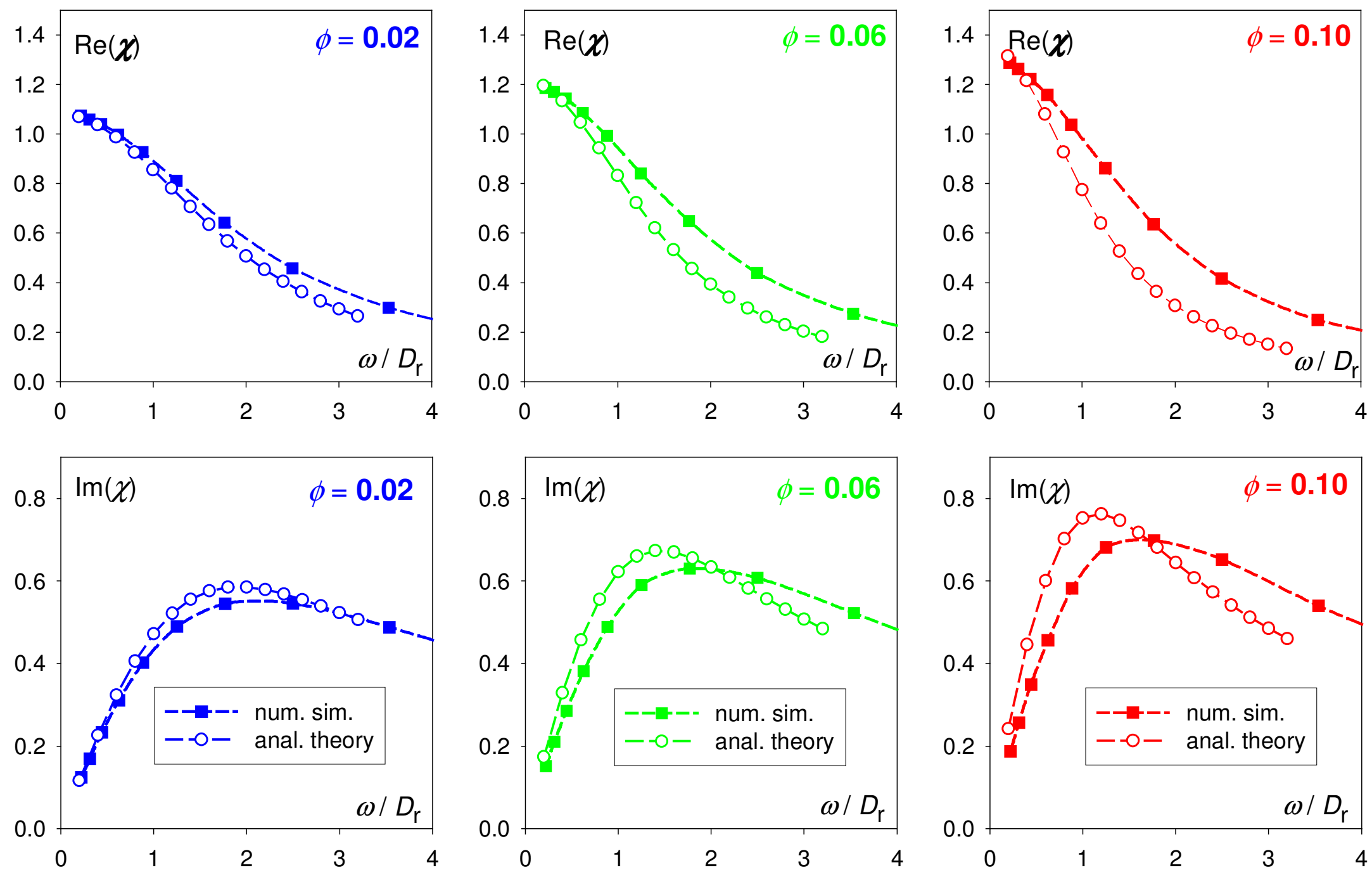

Fig. 6